\documentclass[twocolumn,pra,superscriptaddress,showpacs,amsmath]{revtex4}

\usepackage{graphicx}

\begin{document}

\title
     {Coherent interaction of laser pulses in a resonant optically dense
      extended medium under the regime of strong field-matter coupling}
\date{August 28, 2003}

\author{V. S. Egorov}
\author{V. N. Lebedev}
\author{I. B. Mekhov} \email{Mekhov@yahoo.com}
\author{P. V. Moroshkin}
\author{I. A. Chekhonin}
\affiliation{St. Petersburg State University, Department of Optics,
Ulianovskaya 1, Petrodvorets, 198504 St. Petersburg, Russia}
\author{S. N. Bagayev}
\affiliation{Institute of Laser Physics, Siberian Branch of the Russian
Academy of Sciences, Lavrentyeva 13/3, 630090 Novosibirsk, Russia}

\begin{abstract}

Nonstationary pump-probe interaction between short laser pulses propagating
in a resonant optically dense coherent medium is considered. A special
attention is paid to the case, where the density of two-level particles is
high enough that a considerable part of the energy of relatively weak
external laser-fields can be coherently absorbed and reemitted by the medium.
Thus, the field of medium reaction plays a key role in the interaction
processes, which leads to the collective behavior of an atomic ensemble in
the strongly coupled light-matter system. Such behavior results in the fast
excitation interchanges between the field and a medium in the form of the
optical ringing, which is analogous to polariton beating in the solid-state
optics. This collective oscillating response, which can be treated as
successive beats between light wave-packets of different group velocities, is
shown to significantly affect propagation and amplification of the probe
field under its nonlinear interaction with a nearly copropagating pump pulse.
Depending on the probe-pump time delay, the probe transmission spectra show
the appearance of either specific doublet or coherent dip. The widths of
these features are determined by the density-dependent field-matter coupling
coefficient and increase during the propagation. Besides that, the widths of
the coherent features, which appear close to the resonance in the broadband
probe-spectrum, exceed the absorption-line width, since, under the
strong-coupling regime, the frequency of the optical ringing exceeds the rate
of incoherent relaxation. Contrary to the stationary strong-field effects,
the density- and coordinate-dependent transmission spectra of the probe
manifest the importance of the collective oscillations and cannot be obtained
in the framework of the single-atom model.

\end{abstract}


\maketitle

\section{Introduction}

Investigation of the nonlinear medium response to coherent radiation is a
powerful tool to analyze microscopic properties of matter and ultrafast
processes taking place therein. Moreover, detailed study of resonant
light-matter interaction enables the development of new techniques of active
control of processes under investigation. Optically dense media attract
significant attention, since they provide information about phenomena taking
place in macroscopic ensembles of resonant particles. On the other hand, a
dense medium represents a rather complicated object, because it can strongly
modify characteristics of probing radiation by coherent absorption and
reemission of photons. These processes lead to appearance of the medium
reaction field, which plays an essential role and becomes comparable to the
field of an external radiation source. Thus, the coupled electromagnetic
field and optically dense matter should be considered self-consistently as a
single system with specific characteristics and evolution.

In the limit, where the reemission field is negligible in comparison to the
strong, externally applied one, the model of single driven-atom is usually
used. Atomic variables can be calculated on the basis of Bloch equations
without taking into account Maxwell equations. Thus, the spectral features of
the process are fully determined by the strong external field, but not by the
density of the medium. In this context, phenomena such as Rabi-sideband
generation due to the stationary probe-pump Mollow-Boyd effect \cite{boyd} or
to transient Rabi flopping \cite{allen} can be mentioned.

The main peculiarity of the coherent interaction between light and optically
dense media is related to collective behavior of resonant particles due to
their interaction through the reemission field. This behavior can
significantly modify the mutual light-matter dynamics, which cannot be
described in the framework of the single-atom model. In the case, where the
high density of resonant medium enables the atomic system to coherently
absorb and reradiate a considerable part of the energy of the external
electromagnetic field, the fast excitation interchanges between field and
matter appear, which is analogous to polariton beats in the solid-state
optics. The frequency of such collective oscillations is determined by the
density-dependent field-matter coupling coefficient and exceeds the
relaxation rates of a medium. Under the cavity interaction, this
``strong-coupling regime'' results in vacuum Rabi-oscillations
\cite{kaluzny,zhu,rempe,messin,raimond}. In the field of cavity QED,
nonclassical properties of the strong-coupling regime (photon antibunching
and sub-Poissonian statistics \cite{rempe}, squeezing \cite{messin}, and
application for entanglement \cite{raimond}) have been discussed. Under the
free-space pulse propagation, the strong-coupling regime leads to the
formation of the collective optical ringing \cite{crisp,prasad}. The
generation of such a signal may appear even under the weak-field excitation,
without essential changes in the population difference of a medium. This fact
is different from the case of the resonant pulse breakup into $2\pi$ solitons
of self-induced transparency \cite{allen}, where the incident pulse is strong
enough to invert a part of a medium (the area of the external field should be
at least greater than $\pi$). In this context, the optical ringing represents
a nonsolitonic solution in the form of a $0\pi$ pulse.

In this paper, we present a study of coherent free-space interaction between
two (probe and pump) short laser pulses propagating in a two-level extended
medium under the conditions of the strong-coupling regime. We center our
discussion on the question, how the collective oscillatory response of an
optically dense medium, excited by a short pump pulse, affects the
transmission spectrum of the probe field. Since the optical ringing
originates from the linear response of a dense resonant medium, we consider
the case of a weak pump field, which does not invert atoms. Hence, all
obtained transformations of the probe pulse, particularly its amplification,
take place in a medium without population inversion.

The optical ringing represents an essentially transient phenomenon and can be
observed in a time scale shorter than the relaxation times of a medium. Due
to its general character, such oscillating response accompanies a variety of
coherent phenomena taking place under the short pulse propagation
\cite{crisp} in dense media, and should be taken into account in analysis of
experimental data. Having a superradiant character, it was studied in the
contexts of the so-called ``polarium model'' \cite{prasad} and oscillatory
regime of Dicke superradiance \cite{macgil}. Its influence on the quantum
coherent control and pulse shaping techniques was mentioned in
Refs.~\cite{sweetser,felinto00,dudovich}. The works \cite{laenen} were
devoted to the determination of relaxation times of homogeneously and
inhomogeneously broadened spectral lines of molecular substances in the
presence of the optical ringing. In Refs.~\cite{sweetser,kallman,felinto00},
different aspects of the coherent interaction of ultrashort laser pulses with
extended multilevel atomic media, particularly two-photon absorption, were
considered. The existence of ultrafast collective oscillations were
demonstrated in Ref.~\cite{matusovskyJOSA96}, where not only the external
laser pulse, but also the optical ringing itself belonged to the femtosecond
time scale.

Besides the atomic and molecular media, the optical ringing was also
rediscovered for the case of solid state exciton-polaritons \cite{frohlich}.
It was treated as quantum beats between two branches of polariton dispersion
curve corresponding to a single medium resonance, contrasting to the case of
usual quantum beats between different optical transitions. Moreover, the
transition from the strong-field regime of Rabi flopping to the weak-field
optical ringing under the strong-coupling regime was recently observed in
Ref.~\cite{nielsen}. The femtosecond ringing was obtained for the cases of
both bulk crystals \cite{nusse} and multiple quantum well structures
\cite{kim}.

In the presence of the optical ringing, the interaction of two intersected
probe and pump beams (this configuration is also used in the present work)
was considered for the cases of atomic media \cite{kinrot,weisman} and solids
\cite{kim,pantke,schillak,bakker}. Most of the results of these studies are
related to the time-integrated and time-resolved properties of the diffracted
four-wave mixing signal. In contrast, our discussion is centered on the
modification of the probe field, and particularly, on its spectral
properties. In our previous works
\cite{vasilyev94,bagayev02,bagayev03,bagayevPRA}, we have studied both
theoretically and experimentally interaction of broadband polychromatic
pulses of a multimode dye-laser without mode locking in an optically dense
resonant medium (metastable neon atoms in gas discharges). Different spectral
types of probe amplification were obtained and analyzed. Our present paper is
devoted to the interaction of short transform-limited pulses, since
fundamentals of the phenomena have been considered arise from the coherent
interaction between short pulses and a dense medium.

The paper is organized as follows. In Sec.~II, the main equations of the
theoretical model are obtained. In Sec.~III, we outline some specific
features of the collective optical ringing originating from the dispersion of
a dense resonant medium. The numerical results and discussion of the
probe-field propagation in the presence of a pump pulse are presented in
Sec.~IV. Main results of the paper are summarized in Sec.~V.

\section{Equations of the model}

The theory of transient processes of interaction between electromagnetic
field and resonant media is based on the joint solution of the semiclassical
Maxwell-Bloch equations \cite{allen}. We restrict our study to the two-level
system approximation assuming that interaction between the field and
neighboring atomic transitions is negligibly small. The electric field
$E(t,{\bf r})$ and the polarization of the medium $P(t,{\bf r})$ can be
written in the form

\begin{eqnarray*}
E(t,{\bf r})=\frac{1}{2}\frac{\hbar}{d}\left(\Omega_s(t,{\bf r})e^{i\omega_0
t}+
\text{c.c.}\right), \\
P(t,{\bf r})=\frac{1}{2} nd\left(p_s(t,{\bf r})e^{i\omega_0 t-i\pi/2}+
\text{c.c.}\right),
\end{eqnarray*}
introducing the complex amplitudes $\Omega_s(t,{\bf r})$ and $p_s(t,{\bf r})$
slowly varying in time but having arbitrary spatial dependence. Here, $d$ and
$\omega_0$ are the electric dipole moment and the resonance frequency of the
atomic transition, $n$ is the density of atoms. The amplitude
$\Omega_s(t,{\bf r})$ is expressed in the units of the Rabi frequency of the
electric field.

Using the rotating wave approximation, the system of Bloch equations can be
written as

\begin{subequations}\label{1}
\begin{eqnarray}
\frac{\partial p_s}{\partial t} & = & \Omega_s D -\gamma_2 p_s, \\
\frac{\partial D}{\partial t} & = & -\frac{1}{2}\left(\Omega_s p_s^* +
\Omega_s^* p_s\right)- \gamma_1\left(D-D^{eq}\right),
\end{eqnarray}
\end{subequations}
where $D$ is the population difference of an atom, $D^{eq}$ is the value of
$D$ in the absence of external field (the value $D=1$ corresponds to an atom
in the ground state), and $\gamma_1$ and $\gamma_2$ are the relaxation rates.

The problem of the interaction between two intersected plane linearly
polarized waves is considered. In the case of nearly copropagating pulses,
the amplitude of the field is given by

\begin{equation}\label{2}
\Omega_s=\Omega_0(t,z)e^{-ik_{0}z}+\Omega_1(t,z)e^{-i{\bf k}_{1}{\bf r}},
\end{equation}
with $\Omega_0$ and $\Omega_1$ slowly varying in space. The field $\Omega_0$
with the wave vector ${\bf k}_0$, which is parallel to $z$ axis, corresponds
to the strong pump wave, whereas $\Omega_1$ with ${\bf k}_1$ wave vector is
assumed to be a weak probe field propagating at the small angle $\varphi$
with respect to $z$ direction. Nonlinear interaction of the intersected waves
leads to the appearance of spatial polarization harmonics with the wave
vectors ${\bf k}_0+m\Delta{\bf k}$ ($m=0, \pm 1, \pm 2,...$, $\Delta{\bf
k}={\bf k}_1-{\bf k}_0$) and harmonics of the population difference with
$m\Delta{\bf k}$ wave vectors:

\begin{subequations}\label{3}
\begin{eqnarray}
p_s=\sum_{m=-\infty}^{\infty}p_m(t,z)e^{-i({\bf k}_{0}{\bf r}+m\Delta{\bf
k}{\bf r})}, \label{3a}
\\
D=\sum_{m=-\infty}^{\infty}D_m(t,z)e^{-im\Delta{\bf k}{\bf r}} , \qquad
D_m=D_{-m}^* . \label{3b}
\end{eqnarray}
\end{subequations}
The emission of $p_0$ and $p_1$ polarizations corresponds to the pump and
probe fields, respectively. The emission of higher harmonics is considered to
be suppressed in a thick medium, due to mismatch in dispersion relation.
Substituting the expansions (\ref{2}) and (\ref{3}) into Bloch equations
(\ref{1}), adding Maxwell equations, and using the first-order perturbation
theory in respect of the small amplitude of the probe field, one can get the
coupled Maxwell-Bloch system, describing propagation of the strong pump
field,

\begin{subequations}\label{4}
\begin{eqnarray}\
c\frac{\partial \Omega_0}{\partial z}&+&\frac{\partial \Omega_0}{\partial
t}=-\omega_c^2 p_0 , \label{4a}\\
\frac{\partial p_0}{\partial t}&=&\Omega_0
D_0-\gamma_2 p_0,  \label{4b}\\
\frac{\partial D_0}{\partial t}&=&-\frac{1}{2}\left(\Omega_0 p_0^* +
\Omega_0^* p_0\right)- \gamma_1\left(D_0-D^{eq}\right), \label{4c}
\end{eqnarray}
\end{subequations}
and a weak probe,

\begin{subequations}\label{5}
\begin{eqnarray}
c \cos\varphi\frac{\partial \Omega_1}{\partial z}+\frac{\partial
\Omega_1}{\partial t}=-\omega_c^2 p_1 , \label{5a}\\
\frac{\partial p_1}{\partial t}=\Omega_1 D_0+\Omega_0 D_1-\gamma_2 p_1, \label{5b}\\
\frac{\partial D_1}{\partial t}=-\frac{1}{2}\left(\Omega_1 p_0^* +
\Omega_0^* p_1+\Omega_0 p_{-1}^*\right)- \gamma_1 D_1, \label{5c}\\
\frac{\partial p_{-1}^*}{\partial t}=\Omega_0^* D_1-\gamma_2 p_{-1}^*,
\label{5d}
\end{eqnarray}
\end{subequations}
where

\begin{equation}\label{6}
\omega_c=\sqrt{\frac{2\pi d^2 \omega_0 n}{\hbar}}
\end{equation}
is the cooperative frequency of the medium, which plays a role of the
coupling coefficient between field and matter. We will present the numerical
results of our work in the dimensionless form in the units of $\omega_c$.

If the functions $\Omega_{0,1}(t,z)$ are real (exactly resonant interaction
without phase modulation of input pulses), the areas of the pulses, defined
as

\begin{equation*}
s_{0,1}(z)=\int_{0}^{\infty}\Omega_{0,1}(t,z)dt ,
\end{equation*}
are very important parameters characterizing coherent propagation of short
laser pulses. In the case of complex functions $\Omega_{0,1}(t,z)$, the
definition of the area does not have a general form.

\section{Peculiarities of the optical ringing in a single laser-beam}

Contrasting to the case of nonlinear pulse breakup into $2\pi$ solitons
\cite{allen}, the origin of collective oscillations can be traced to the
linear light-matter interaction. In this section, we outline main features of
this phenomenon, considering the propagation of the single (pump) field
described by Eqs. (\ref{4}). In the linear limit of a weak (small area)
field, where the atoms are assumed to be in the ground state with $D(t,z)=1$,
Eqs. (\ref{4a}) and (\ref{4b}) can be written as a single second-order
differential equation

\begin{equation}\label{7}
c\frac{\partial^2 \Omega}{\partial z \partial t }+\frac{\partial^2
\Omega}{\partial^2 t}+\gamma_2\left( c\frac{\partial \Omega}{\partial
z}+\frac{\partial \Omega}{\partial t}\right) +\omega^2_c\Omega=0 ,
\end{equation}
describing oscillations of the laser field, which propagates in a medium with
high cooperative frequency $\omega_c$ (\ref{6}). In this section, considering
the propagation of a single beam, we have omitted the subscript of the field
amplitude. The problem of the free-space propagation of a short laser pulse
entering an initially unperturbed medium can be solved taking the time
Fourier transform of Eq. (\ref{7}). The inverse Fourier transform then gives
the solution in the following form (cf. Ref.~\cite{crisp}):

\begin{equation}\label{8}
\Omega(t,z)=\frac{1}{2\pi}\int_{-\infty}^{\infty}F_\text{in}(\omega)
e^{-ik(\omega)z}e^{i\omega t}d\omega ,
\end{equation}
where $F_\text{in}(\omega)$ is the spectrum of the laser pulse at the input
of the medium at $z=0$. The spatial evolution of spectral components is
determined by the wave vector $k(\omega)$:

\begin{equation}\label{9}
ck(\omega)=\omega-\frac{\omega_c^2}{\omega-i\gamma_2}.
\end{equation}
Here, the frequency $\omega$ and the wave vector $k$ are defined as detunings
from the resonant values $\omega_0$ and $\omega_0 /c$, respectively. The real
part of $k(\omega)$ determines dispersive characteristics of propagating
field, whereas its imaginary part determines the absorption.

\begin{figure}
\includegraphics{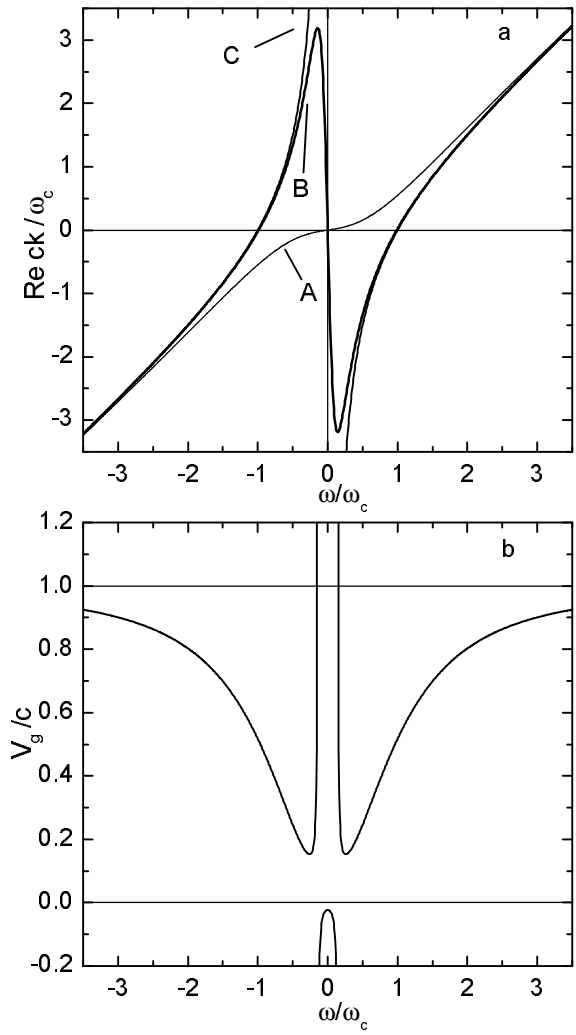}
\caption{\label{fig1}Dispersion characteristics of dense resonant medium. (a)
linear dispersion $\text{Re} k(\omega)$ for $\gamma_2/\omega_c=1.1$ (curve
A), 0.15 (curve B), and 0.0 (curve C); (b) group velocity $V_g(\omega)$ for
$\gamma_2/\omega_c=0.15$.}
\end{figure}

Depending on the relation between the field-matter coupling coefficient
$\omega_c$ and the rate of incoherent relaxation $\gamma_2$, two
qualitatively different frequency dependencies of the real part of
$k(\omega)$ are possible, which is demonstrated in Fig.~\ref{fig1}(a). If the
relaxation rate exceeds the cooperative frequency, only one value of the
frequency $\omega$ corresponds to each wave vector $k$ (curve A). In the
opposite case, where

\begin{equation}\label{10}
\omega_c > \gamma_2,
\end{equation}
there may exist three frequencies, that correspond to the same value of $k$
(curve B). The central frequency falls to the region of the anomalous
dispersion with strong absorption of the field, whereas two other solutions
appear at the wings of the spectral line, where absorption may be very small.

The existence of photons of equal wave vectors, but of different frequencies
is a characteristic property of the resonant light-matter interaction in
optically dense coherent media. This phenomenon can be treated as a
density-dependent splitting of normal modes in the system of strongly coupled
field-matter oscillators (polaritons). In the limiting case of purely
coherent interaction ($\gamma_2=0$), the polariton dispersion is represented
by two separated anticrossing branches, displayed by the curve C in
Fig.~\ref{fig1}(a). The coherent beating between normal modes of these kind
can be observed, if the spatial spectrum of the field is fixed, for example,
by initial conditions or by the presence of a cavity. In this context,
effects such as cavity collective vacuum Rabi-oscillations \cite{kaluzny,zhu}
and oscillatory regime of superradiance under side-excitation of a sample
\cite{macgil} can be mentioned. In these cases, fixing of wave vectors leads
to the appearance of a characteristic doublet in the field spectrum, which
corresponds to the normal modes and manifests the existence of the
``strong-coupling regime'' determined by Eq. (\ref{10}).

In the situation under analysis (\ref{7}), (\ref{8}), where a laser pulse
propagates in an initially unperturbed medium, the frequency content of the
output field is entirely determined by the spectrum of the external source
$F_\text{in}(\omega)$. Due to the linearity of interaction, the absolute
value of the output spectrum does not vary during the propagation, except for
the trivial appearance of the absorption line. Thus, in this case, contrary
to collective vacuum Rabi-oscillations, no coherent density-dependent
spectral features can be extracted from spectral measurements. At the same
time, coherent oscillations can be observed under analysis of the temporal
behavior.

Using the theorem about Fourier transform of a product, the solution
(\ref{8}) can be rewritten in the following form (cf. Ref.~\cite{laenen}):

\begin{eqnarray}\label{11}
\Omega(t,z)=\Omega_\text{in}(\tau) \nonumber\\
-\int_{0}^{\tau} \Omega_\text{in}(\tau-t') e^{-\gamma_2
t'}\omega_c\sqrt{\frac{z}{ct'}}J_1 (2\omega_c\sqrt{zt'/c})dt'
\end{eqnarray}
for $\tau \ge 0$, and $\Omega(t,z)=0$ for $\tau < 0$. Here $\tau=t-z/c$,
$\Omega_\text{in}(t)$ is the complex amplitude of the electric field at the
input of the medium, and $J_1(x)$ is the first-order Bessel function. The
integral kernel represents the oscillatory response of the atomic ensemble to
the short delta-pulse [$\Omega_\text{in}(t)=\delta(t)$], and thus represents
the Green function of the problem.

The optical ringing, which is described by the Bessel function in
Eq.~(\ref{11}), corresponds to oscillations with frequency varying in time
and space. This fact reveals the principal role of propagation effects in the
formation of the optical ringing. In a cross-section of the medium with the
coordinate $z$, the inverse duration of the first short pulse, which was
reradiated by the medium, is proportional to the quantity

\begin{equation}\label{12}
\omega_D = \frac{\omega_c^2 z}{c} .
\end{equation}
The frequency of subsequent oscillations decreases in time and at the tail of
the ringing is given by $\omega_c\sqrt{z/(c\tau)}$. Hence, to observe at
least the initial stage of the the coherent optical ringing, the following
criterion should be fulfilled:

\begin{equation}\label{13}
\omega_D = \frac{\omega_c^2 z}{c} \gg \gamma_2 .
\end{equation}
The frequency of the optical ringing in its initial stage increases with
propagation distance $z$ and the medium density $n$. Thus, it is proportional
to the number of atoms interacting via the reemission field (here, only
propagation in the forward direction is considered). Such a dependence
directly demonstrates the collective character of the field reemission by the
atomic ensemble. The quantity $\omega_D$ (\ref{12}) is a well-known parameter
in the theory of Dicke superradiance \cite{prasad,macgil}. Unlike a single
superradiant pulse, which is considered as fast collective relaxation of
atoms to the ground state, the optical ringing represents fast collective
oscillations that nevertheless remain coherent in a time scale of
$1/\gamma_2$, which may be much longer than the duration of a single pulse of
superradiance.

The condition of the appearance of the optical ringing (\ref{13}) is
different from that of the collective vacuum Rabi-oscillations (\ref{10}).
Nevertheless, the relation between the field-matter coupling coefficient
$\omega_c$ (\ref{6}) and the rate of incoherent relaxation of the medium is
very important for both types of collective oscillations. The relaxation
rates included in our model may account for radiative and collisional
broadening of a spectral line, while the inhomogeneous Doppler broadening is
not taken into account. Nevertheless, if the processes of interest are much
faster than the relaxation and the detailed shape of a narrow absorption
contour is not of importance, the Doppler broadening may be also
approximately included in the relaxation rates considered \cite{crisp}.

The expressions (\ref{10}) and (\ref{13}) can be considered as conditions
determining a threshold value for the atomic density, which should be
exceeded to obtain collective oscillations. It is important to stress, that
these conditions are essentially weaker than the one necessary for other
density-dependent phenomena in resonant media, such as total backward
reflection \cite{prasad} and local field effects \cite{benaryeh} leading to
the Lorentz-Lorenz frequency shift and intrinsic optical bistability. This
fact substantiates our model presented in Sec.~II, which does not take into
account the latter high-density effects.

The statement, that the initial optical ringing frequency (\ref{12}) displays
the unlimited increase during the propagation, is valid only for the
idealized situation of an input pulse having an infinitely broad spectrum. To
describe more realistic conditions, we will present a simple analytical
expression giving the relation between temporal behavior of the field and the
spectrum of the input pulse.

For this purpose, the solution (\ref{8}) can be analyzed using the stationary
phase method, which gives the following asymptotical value for an integral
expression:

\begin{eqnarray}\label{14}
&&\int_{-\infty}^{\infty}{f(\omega)e^{i\lambda s(\omega)}d\omega} \nonumber\\
&&\sim
\sum_{\omega_s}\sqrt{\frac{2\pi}{\lambda|s''(\omega_s)|}}{f(\omega_s)e^{i\lambda
s(\omega_s)\pm i\pi/4}}
\end{eqnarray}
for $\lambda \rightarrow \infty$ and $s''(\omega_s)\ne 0$. The sum is taken
over all points of stationary phase $\omega_s$ such that $s'(\omega_s)=0$;
the plus or minus sign is chosen for $s''(\omega_s)>0$ and $s''(\omega_s)<0$,
respectively. In the limit of coherent interaction, $\gamma_2 =0$, the
solution (\ref{8}), (\ref{9}) can be reduced to the integral on the left-hand
side of Eq.~(\ref{14}) by the following substitutions:

\begin{equation} \nonumber
f(\omega)=\frac{F_\text{in}(\omega)}{2\pi},\qquad
\lambda=\frac{\omega_cz}{c},\qquad
s(\omega)=\frac{\omega}{\omega_c}\beta+\frac{\omega_c}{\omega},
\end{equation}
where $\beta=c\tau/z$. In this case, there exist two stationary phase points
$\omega_{s1,2}=\pm\omega_g$,

\begin{equation}\label{15}
\omega_{g}=\omega_c \sqrt{\frac{z}{c\tau}}.
\end{equation}
Using Eq.~(\ref{14}), the asymptotic solution of the problem for high values
of $\omega_cz/c$, finite values of the parameter $\beta$, and a smooth
function $F_\text{in}(\omega)$ can be written as

\begin{eqnarray}\label{16}
\Omega(t,z) &\sim&
\frac{1}{2\tau\sqrt{2\pi}}\left(2\omega_c\sqrt{z\tau/c}\right)^{1/2}
\nonumber\\
&\times&\biggl[F_\text{in}\left(\omega_c \sqrt{\frac{z}{c\tau}}\right)
e^{i2\omega_c\sqrt{z\tau/c}+i\pi/4} \nonumber\\
&+&F_\text{in}\left(-\omega_c\sqrt{\frac{z}{c\tau}}\right)
e^{-i2\omega_c\sqrt{z\tau/c}-i\pi/4}\biggr].
\end{eqnarray}

Expression (\ref{16}) shows that the field amplitude $\Omega(t,z)$ is
determined by the complex amplitudes of the input spectrum at the frequencies
$\pm\omega_g$ (\ref{15}). These frequencies are symmetrically placed around
the frequency of the atomic transition and depend on $z$ and $t$. Under the
increase of the propagation distance $z$, the field amplitude is determined
by the spectral components lying farther and farther from the resonance.
Thus, the increase of the ringing frequency is limited by the frequency of
non-negligible spectral components farthest from the resonance.

On the other hand, in a fixed cross-section of the medium, the time evolution
of the field is determined by the spectral components, which are placed
closer and closer to the resonance frequency. In other words,
Eqs.~(\ref{15}), (\ref{16}) show that in the cross-section $z$, a spectral
component of an arbitrary frequency $\omega$ determines the field amplitude
at the time moment $\tau(\omega)$ such that

\begin{equation}\nonumber
\tau(\omega)=\frac{\omega_c^2z}{\omega^2c}.
\end{equation}
One can easily show that $\tau(\omega)$ corresponds to the time delay related
to the difference between the speed of light in vacuum and the group velocity
of a wave-packet of frequency $\omega$. The frequency-dependent group
velocity $V_g(\omega)$ is defined by the derivative of the real part of the
wave vector $k(\omega)$ (\ref{9})

\begin{eqnarray*}
\frac{V_g(\omega)}{c}\equiv \left(c\frac{d \text{Re}
{k}}{d\omega}\right)^{-1}
=1-\frac{\omega_c^2(\omega^2-\gamma_2^2)}{(\omega^2+\gamma_2^2)^2+
\omega_c^2(\omega^2-\gamma_2^2)}
\end{eqnarray*}
and is displayed in Fig.~\ref{fig1}(b). In the case $\omega_c \gg \gamma_2$,
the width of the absorptive (anomalous dispersion) area $2\gamma_2$ is much
smaller than the width of the $V_g(\omega)$ contour, which is equal to
$2\omega_c$ and hence increases with the atomic density. Figure \ref{fig1}(b)
shows, that in an optically dense coherent medium, the group velocity
significantly varies over the spectrum and can be essentially reduced near
the resonance.

Thus, long oscillations of the optical ringing can be treated as successive
beats between light wave-packets of frequencies symmetrical with respect to
the resonance \cite{frohlich}. Equation~(\ref{16}) gives the quantitative
characteristics of this treatment. In contrast to collective vacuum
Rabi-oscillations, which are explained by the beating between two
monochromatic waves of equal wave vectors [see Fig.~\ref{fig1}(a)], the
free-space optical ringing originates from the beats between wave-packets of
equal group velocities, which should be present in the broad spectrum of a
short input pulse [see Fig.~\ref{fig1}(b)].

Moreover, Eq.~(\ref{16}) shows that to observe optical ringing in the
absolute value of the electric field $|\Omega(t,z)|$, the spectrum of the
input pulse $F_\text{in}(\omega)$ should excite both red and blue wings of
the spectral line. This fact confirms the essentially resonant character of
the collective optical ringing. In the case, where the input spectrum covers
only one of the wings, the oscillations in the absolute value are absent,
though they are still present in the quadrature components of the field,
which are determined by the real and imaginary parts of $\Omega(t,z)$.

\begin{figure}
\includegraphics{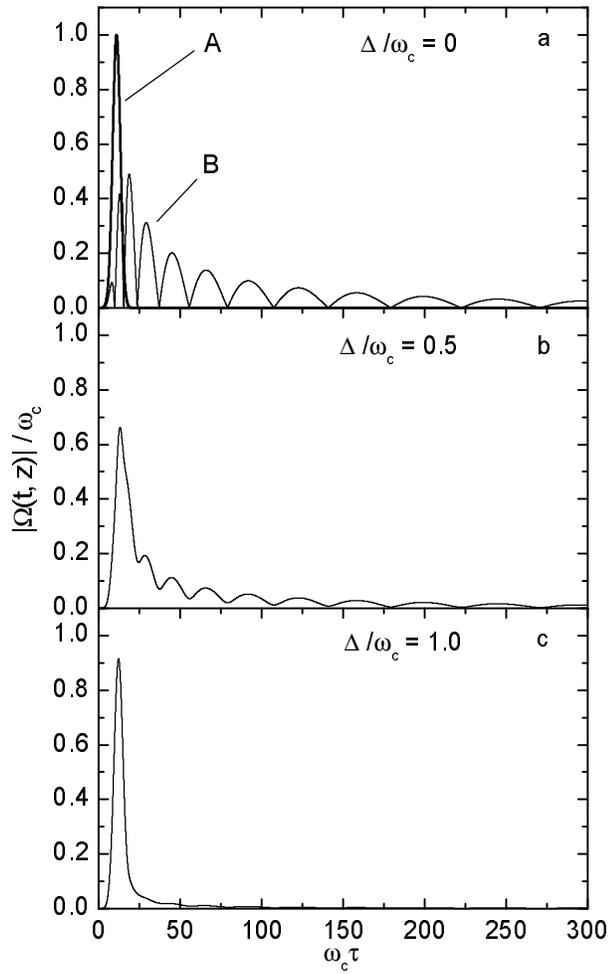}
\caption{\label{fig2}Optical ringing at different detunings of the pulse
spectrum from resonance frequency. (a) input pulse (curve A) and output pulse
for $\Delta/\omega_c=0.0$ (curve B); (b) output pulse for
$\Delta/\omega_c=0.5$; (c) output pulse for $\Delta/\omega_c=1.0$;
$\gamma_2/\omega_c=10^{-3}$, FWHM of the spectrum $\gamma_\text{sp}/\omega_c
= 1.0$, propagation distance $\omega_c z/c=1.0$.}
\end{figure}

Figure \ref{fig2} displays the temporal behavior of optical ringing [the
solution of Eq.~(\ref{7})] excited by a short gaussian-shaped input pulse

\begin{equation}\label{17}
\Omega_\text{in}(t)=\frac{s}{a\sqrt{\pi}}e^{-[(t-t_0)/a]^2+i\Delta t}
\vartheta(t)
\end{equation}
at different detunings $\Delta$ between the pulse central frequency and the
frequency of the resonance. The full width at half maximum (FWHM) of the
spectrum $F_\text{in}(\omega)$, which corresponds to the input pulse
(\ref{17}), is determined through the parameter $a$ and given by

\begin{equation}\label{18}
\gamma_\text{sp}=\frac{4}{a}\sqrt{\ln 2}.
\end{equation}
In Eq.~(\ref{17}), the time delay $t_0 \gg 2\pi/\gamma_\text{sp}$ and the
Heaviside step-function $\vartheta(t)$ are included to fulfill the formal
requirement of $\Omega_\text{in}(t)=0$ for $t<0$, quantity $s$ is the area of
the input pulse.

Figure \ref{fig2} shows that for the fixed parameters of the medium, the
amplitude of the ringing in $|\Omega(t,z)|$ essentially decreases under the
increase of the detuning $\Delta$ and, hence, under the decrease of spectral
amplitudes at one wing of the resonance in comparison with amplitudes at
another one. Deformation of the input pulse also decreases manifesting the
transition to the regime of usual nonresonant pulse propagation with a mean
group velocity. At the same time, the frequency of optical ringing, which
depends on the characteristics of the medium, does not change, since it is
determined by the oscillating exponents in Eq.~(\ref{16}) but not by the
amplitudes of the wave-packets $F_\text{in}(\omega)$. This fact was
experimentally demonstrated, particularly, in
Refs.~\cite{matusovskyJOSA96,nusse}.

\section{Nonlinear pump-probe interaction}

The dynamics of coherent collective phenomena under the regime of strong
light-matter coupling should significantly affect the nonlinear interaction
of laser fields in resonant media. The collective model of the ``spectrum
condensation'' (or ``self-frequency-locking'')
\cite{runge,rubinov,vasilyev94,bertseva} under generation of a multimode dye
laser with an intracavity absorbing cell was proposed in
Ref.~\cite{vasilyev94}. Particularly, in the case of short transform-limited
pulses, the ``spectrum condensation'' was investigated in
Refs.~\cite{runge,rubinov}. The giant parametric amplification of polaritons
was observed in semiconductor microcavities under the large vacuum
Rabi-splitting \cite{savvidis,messin,huynh}.

In this paper, we present the first investigation of the influence of the
collective optical ringing on the transmission of a probe pulse in the
presence of a short pump pulse. In the presence of the optical ringing, the
configuration of two intersected beams was also considered in investigations
of the resonant four-wave mixing in atomic \cite{kinrot,weisman} and
solid-state media \cite{kim,pantke,schillak,bakker}. In these works, the
collective oscillations of the field were obtained in the time-resolved and
time-integrated studies of a new diffracted pulse. In contrast, our present
paper deals with spectral properties of the probe beam. In
Refs.~\cite{kinrot,kim,pantke,schillak,bakker}, the probe field was not
affected by the pump pulse, since the latter one was considered to be very
weak. Theoretical model of Ref.~\cite{weisman} accounted for the modification
of the probe field, but the carrier frequency of the pump pulse was chosen to
be shifted from the resonance; thus no optical ringing was produced in the
absolute value of the pump field and in the population difference of the
medium. In the present paper, the case of a resonant pump pulse is analyzed,
and the optical ringing is able to affect the probe field also through the
collective oscillations of the population difference $D_0(t,z)$, which
couples two beams intersected at a small angle [cf. Eqs.~(\ref{5a}),
(\ref{5b})]. In Ref.~\cite{czub}, where the interaction of two short resonant
pulses was considered under the exactly collinear configuration, it was
impossible to distinguish between the probe and pump fields, when the long
oscillating field of the medium response appeared.

In our previous works \cite{vasilyev94,bagayev02,bagayev03,bagayevPRA}, we
studied, both experimentally and theoretically, pump-probe interaction of
broadband polychromatic pulses produced by dye lasers without mode locking.
We obtained and analyzed different spectral types of probe amplification in
an optically dense resonant medium. Even in the case of quasistochastic
pulses of correlation time determined by the width of the laser spectrum, the
collective optical ringing was shown to accompany the propagation and
amplification processes. In the experiments, a neon gas discharge containing
significant amount of neon metastable atoms (up to $10^{13}$~cm$^{-3}$) was
used as an optically dense medium. Measurements were carried out at several
neon transitions with the metastable level $2p^53s$ $^3P_2$ as a lower state.
Particularly, for the strongest transition of the wavelength 640.2~nm, the
cooperative frequency $\omega_c/2\pi$ reached a value of 30~GHz, which was
enough to fulfill the conditions of the strong-coupling regime. The widths of
laser spectra greatly exceeded both homogeneous and inhomogeneous broadening
of the spectral line, and took a value of about 200~GHz (FWHM) giving the
correlation time (duration of a single spike in quasistochastic signal) of
about 5~ps.

In contrast to Refs.~\cite{vasilyev94,bagayev02,bagayev03,bagayevPRA}, the
present paper deals with smooth transform-limited pulses, since fundamentals
of the phenomena, have been considered earlier, arise from the coherent
interaction between short pulses and a dense medium. Under the mode locking
of dye lasers, our experimental parameters described above can be used for
quantitative analysis of the results obtained in this work, which are
presented in the dimensionless form. Moreover, as mentioned in the
Introduction, the collective optical ringing was already observed in the
femtosecond time scale. Thus the results obtained has a direct relation to
the study of the so-called ``sharp-line limit'' of resonant light-matter
interaction. In the framework of this limit, the single-pulse propagation was
considered in Refs.~\cite{mikl,ranka,schupper,matusovskyJOSA96}.

The numerical model is based on the solution of the nonlinear system of
Eqs.~(\ref{4}), (\ref{5}) describing the propagation of two intersected
waves. In this system, the weak probe field $\Omega_1(t,z)$ is treated in the
first order of perturbation theory, while the amplitude of the pump field
$\Omega_0(t,z)$ can take arbitrary values. Nevertheless, since the optical
ringing originates from the linear response of a medium under small changes
in the population difference, we center our discussion on the case where the
input area of the pump field takes values less than $\pi/2$. Thus, in the
situation considered, the pump field does not invert a medium. Moreover, the
long ringing can be obtained after both the probe and pump pulses, which
provides interaction of two fields to be much longer than the durations of
the short input pulses.

Equation (\ref{5b}) shows the influence of the pump field on the polarization
component $p_1$ radiating in the direction of the probe $\Omega_1$, whose
propagation is described by Eq.~(\ref{5a}). The first term on the right-hand
side of Eq.~(\ref{5b}) can be treated as forward scattering of the probe wave
$\Omega_1$ on the spatiotemporal oscillations of the population difference
component $D_0(t, z)$, which is determined by the transient collective
ringing in the pump field $\Omega_0$ [cf. Eq.~(\ref{4c})]. The second term in
Eq.~(\ref{5b}) can be described as scattering of the pump field $\Omega_0$ on
the nonstationary  grating of the population difference $D_1(t, z)$ with the
wave vector $\Delta {\bf k}$ (\ref{3b}), which originates from the
spatiotemporal beating between pump and probe waves [cf. Eq.~(\ref{5c})].

In the numerical simulations, the pump pulse at the input of the medium was
chosen in the form (\ref{17}) with the pulse area $s_0 < \pi/2$ and the
spectral FWHM (\ref{18}) $\gamma_\text{sp0}/ \omega_c=10.0$. To provide the
maximum amplitude of pump-matter energy exchange $D_0$, which is determined
by the amplitude of the ringing in $|\Omega_0|$, the pump central frequency
was tuned exactly to the resonance, giving $\Delta_0=0$.

The input probe-pulse was also chosen in the form (\ref{17}) with introducing
an additional time delay $\tau_0$ ($\tau_0<0$ corresponds to the probe pulse
preceding the pump). The probe has different values of $\gamma_\text{sp1}$
and $\Delta_1$, and the area $s_1$ much less than $s_0$ (the exact value of
$s_1$ is not of importance due to the linearity of the equations with respect
to the probe amplitude). The value of the small angle $\varphi$ did not
significantly affect the results and was set to $1^\circ$.

\begin{figure}
\includegraphics{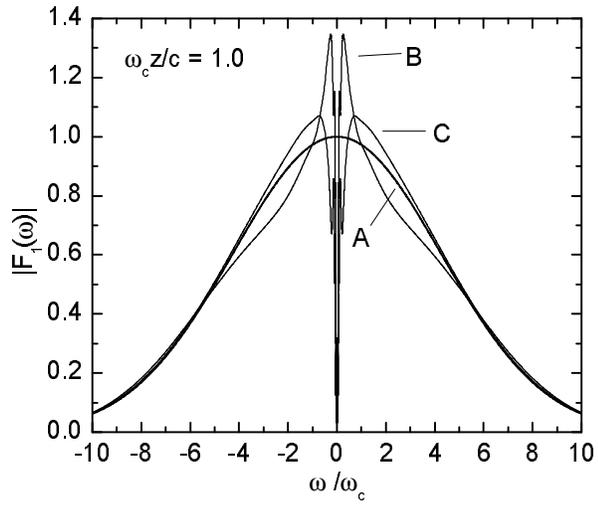}
\caption{\label{fig3}Spectra of the probe field at the input (curve A) and
output of the medium at probe-pump time delays $\omega_c\tau_0=-0.5$ (curve
B) and $\omega_c\tau_0=0.5$ (curve C); $\gamma_{1,2}/\omega_c=10^{-3}$, FWHM
of the probe and pump spectra $\gamma_\text{sp0}/\omega_c =\gamma_\text{sp1}
/ \omega_c = 10$, pump area $s_0=0.49\pi$, propagation distance $\omega_c
z/c=1.0$.}
\end{figure}

A typical result of numerical simulations for the case of exactly resonant
broadband probe pulse (with parameters similar to those of the pump:
$\gamma_\text{sp1}/ \omega_c=10.0$, $\Delta_1=0$) is presented in
Fig.~\ref{fig3}. It shows the spectra of the probe pulse $|F_1(\omega)|$ at
the input of the medium and after the propagation for the cases of the
positive and negative time delays $\tau_0$ between the probe and pump pulses.
The characteristic peculiarity of the output spectra obtained consists in the
fact that, in addition to the narrow absorption line, either broad spectral
doublet for $\tau_0<0$ or dip for $\tau_0>0$ appears. Both of these spectral
features appear close to the resonance. Besides that, at the wings of the
spectra, there exist dumping for $\tau_0<0$ and amplification for $\tau_0>0$.
The integration over the spectrum squared shows that such amplification or
dumping at the spectral wings gives the increase or decrease, respectively,
in the total energy of the pulse under propagation. This fact is a
consequence of the existence of energy exchange between probe and pump, which
interact by the ringing tails originating from the field of medium
reemission.

\begin{figure}
\includegraphics{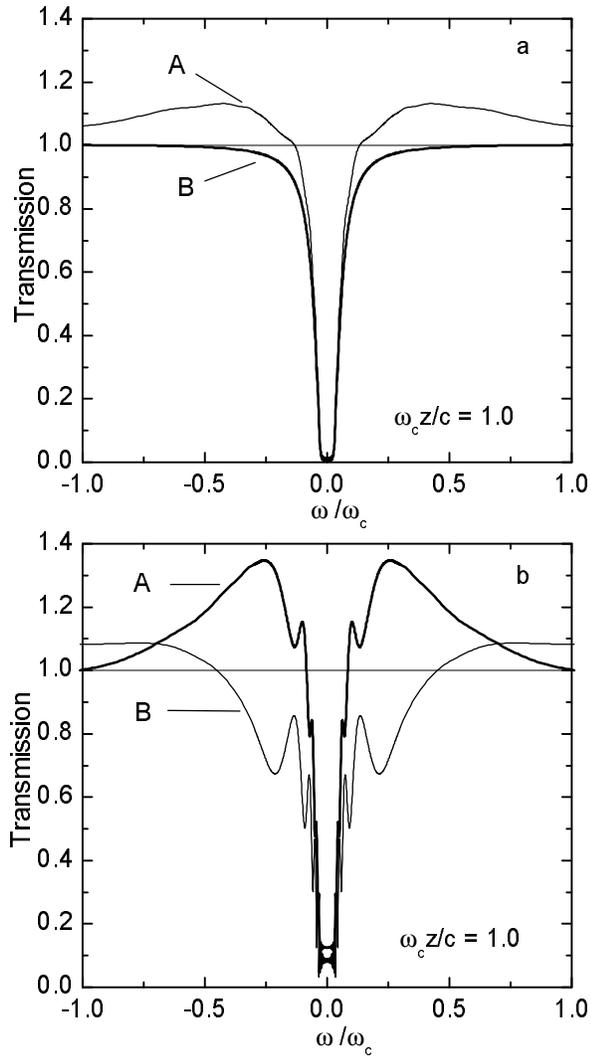}
\caption{\label{fig4}Transmission spectra of the probe field in the vicinity
of resonance for propagation distance $\omega_c z/c=1.0$ and different
probe-pump time delays. (a) $\omega_c\tau_0 =0$ (curve A) and linear
transmission contour (curve B); (b) $\omega_c\tau_0=-0.5$ (curve A) and
$\omega_c\tau_0 = 0.5$ (curve B); $\gamma_{1,2}/\omega_c=10^{-3}$, FWHM of
the probe and pump spectra $\gamma_\text{sp0}/\omega_c =\gamma_\text{sp1} /
\omega_c = 10$, pump area $s_0=0.49\pi$.}
\end{figure}

\begin{figure}
\includegraphics{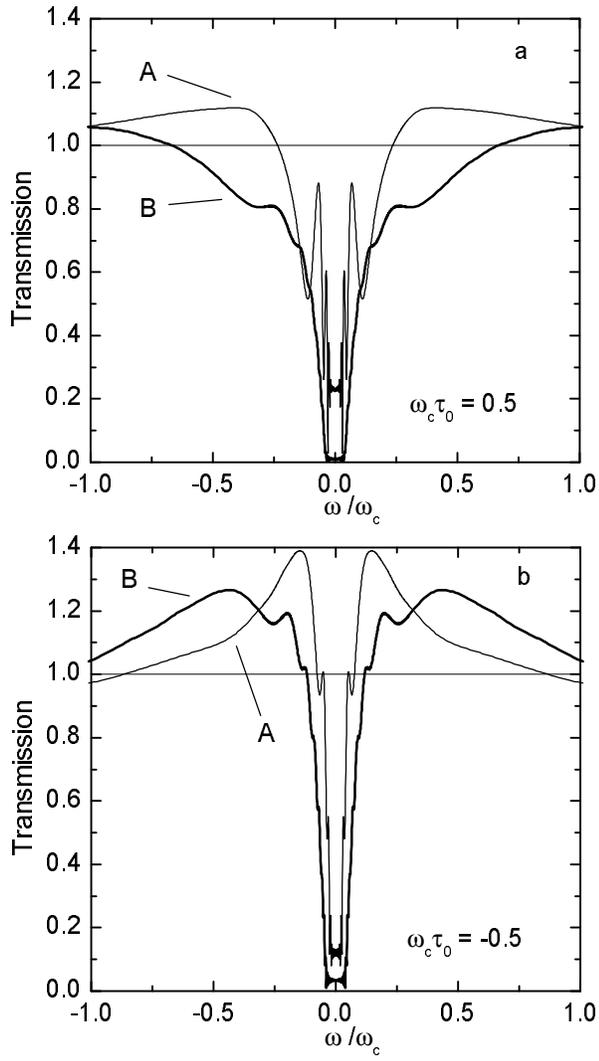}
\caption{\label{fig5}Transmission spectra of the probe field in the vicinity
of resonance for different propagation distances and probe-pump time delays.
(a) $\omega_c\tau_0 = 0.5$ for $\omega_c z/c=0.5$ (curve A) and $\omega_c
z/c=2.0$ (curve B); (b) $\omega_c\tau_0 = -0.5$ for $\omega_c z/c=0.5$ (curve
A) and $\omega_c z/c=2.0$ (curve B); $\gamma_{1,2}/\omega_c=10^{-3}$, FWHM of
the probe and pump spectra $\gamma_\text{sp0}/\omega_c =\gamma_\text{sp1} /
\omega_c = 10$, pump area $s_0=0.49\pi$.}
\end{figure}

The details of the probe-field transmission spectra (output to input spectrum
ratios) in the vicinity of the resonance are presented in Figs.~\ref{fig4}
and~\ref{fig5}.

Figure \ref{fig4} displays the spectra for the dimensionless length of the
medium $\omega_c z/c=1.0$ and different time delays $\tau_0$ between probe
and pump pulses. Figure \ref{fig4}(a) illustrates the transmission of the
probe in the presence of the pump for the probe-pump time delay $\tau_0=0$
(curve A). It also shows the linear transmission contour in the absence of
the pump pulse (curve B). In Fig.~\ref{fig4}(b), the probe spectra with
resonant doublet for $\omega_c \tau_0=-0.5$ and broad dip for $\omega_c
\tau_0= 0.5$ are shown, presenting the detailed description of
Fig.~\ref{fig3}. The comparison of Figs.~\ref{fig4}(a) and \ref{fig4}(b)
shows that the widths of both the spectral doublet and dip are greater than
the width of the linear transmission contour, which corresponds to the
incoherent absorption. Besides that, the broad features are not significant
at $\tau_0=0$. The latter fact demonstrates the importance of the phase
shift, which is introduced by the presence of the time delay between probe
and pump pulses.

Figure~\ref{fig5} compares the probe transmission spectra for different
propagation distances ($\omega_c z/c=0.5$ and 2.0). Figure~\ref{fig5}(a)
presents this comparison for the time delay $\tau_0>0$, whereas the
comparison for $\tau_0<0$ is displayed in Fig.~\ref{fig5}(b). These figures
together with Fig.~\ref{fig4}(b) for $\omega_c z/c=1.0$ show that the widths
of the spectral dip and doublet significantly depend on the coordinate $z$
and increase during the propagation. Such behavior of the spectral features
in the probe field and the fact, that its width exceeds the absorption-line
width, give evidence of the coherent optical ringing importance in the
process of nonlinear interaction.

\begin{figure}
\includegraphics{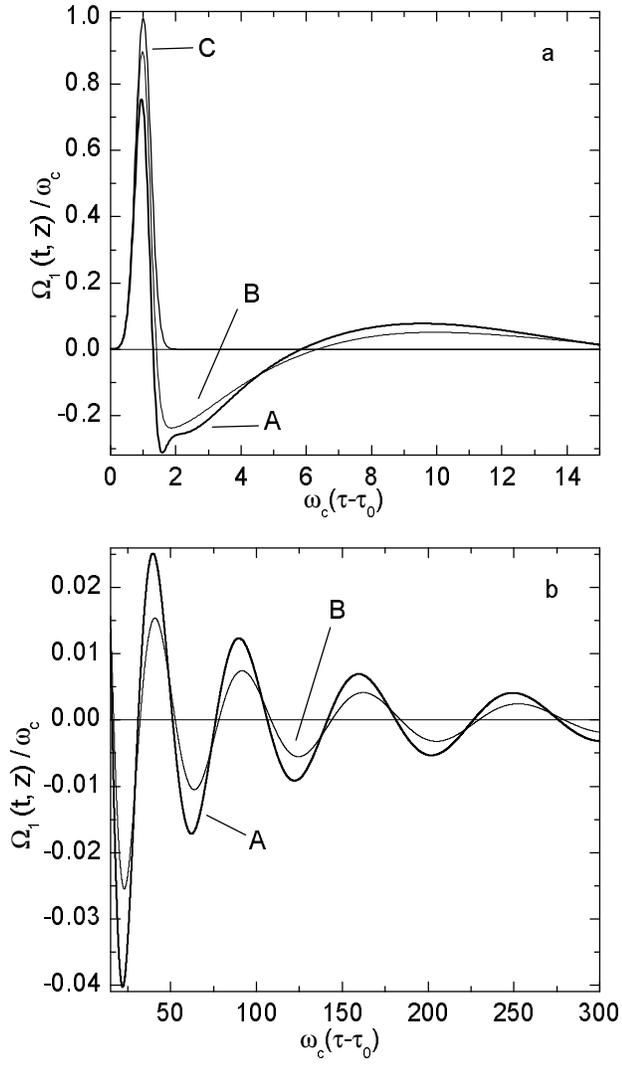}
\caption{\label{fig6}Temporal behavior of the probe field for propagation
distance $\omega_c z/c=1.0$ and probe-pump time delays $\omega_c\tau_0=-0.5$
(curves A) and $\omega_c\tau_0=0.5$ (curves B). (a) initial stage of the
process and the input pulse (curve C); (b) collective oscillating tail of the
probe field; $\gamma_{1,2}/\omega_c=10^{-3}$, FWHM of the probe and pump
spectra $\gamma_\text{sp0}/\omega_c =\gamma_\text{sp1} / \omega_c = 10$, pump
area $s_0=0.49\pi$. Corresponding spectra are shown in Figs.~\ref{fig3}
and~\ref{fig4}(b).}
\end{figure}

Temporal behavior of the probe pulse in the presence of the pump field is
shown in Fig.~\ref{fig6} for $\tau_0<0$ (curves A) and $\tau_0>0$ (curves B).
Here, the argument $\tau$ was shifted to $\tau-\tau_0$ for the convenient
comparison of these curves. The initial stage of the process
[$\omega_c(\tau-\tau_0) < 15$] and the probe pulse at the input of the medium
(curve C) are presented in Fig.~\ref{fig6}(a). The most interesting
contribution to the output signal is given by the coherent oscillating tail
in the form of the collective optical ringing, which is displayed in
Fig.~\ref{fig6}(b) [$\omega_c(\tau-\tau_0) > 15$]. The temporal dynamics
corresponds to the spectra displayed in Figs.~\ref{fig3} and~\ref{fig4}(b).
The dumped oscillations at the tail of the optical ringing (curve B)
correspond to the broad dip in the probe spectrum, while the amplified
oscillations (curve A) reflect the appearance of the spectral doublet.

It is important to stress, that in the problem considered in Sec.~III, due to
the linearity of interaction, no features corresponding to the coherent
optical ringing can be observed in the transmission spectrum, except for the
usual incoherent absorption contour [cf. Fig.~\ref{fig4}(a)]. In contrast,
the results presented in this section show the existence of such spectral
features under the nonlinear pump-probe interaction. The broadening of the
coherent doublet and dip during the propagation is a consequence of the
growing ringing frequency, which increases with the increase in number of
atoms participating in the collective interaction. In other words, the
broadening of the coherent doublet and dip reflects the fact, that
wave-packets of increasing detunings from the resonance being involved in the
creation of the ringing [see Eq.~(\ref{16})], and hence in the dynamical
interaction of two fields.

The relaxation rates $\gamma_1$, $\gamma_2$ and the ``optical density''
$\alpha_0z \equiv \omega_c^2z/(\gamma_2c)$ of a medium, which together
describe the absorption contour, are not the characteristic parameters of the
optical ringing, since they include incoherent parameters of a medium. The
characteristic frequencies of the coherent collective oscillations
(\ref{13}), (\ref{15}) do not depend on the rates of incoherent relaxation
and increase during the propagation. The width of the absorption contour also
increases during the propagation. Nevertheless, since it is determined by the
relaxation rate, it may be significantly smaller than the width of the
coherent features.

The propagation of the probe pulse $\Omega_1(t,z)$ can be treated as a
parametric process under the modulation of parameters in Eqs.~(\ref{5}) by
the pump field. The detailed analysis shows that it is the first term on the
right-hand side of Eq.~(\ref{5b}) with oscillating $D_0(t,z)$ that mainly
contributes to the appearance of the near-resonant features presented in
Figs.~\ref{fig4} and~\ref{fig5}. Thus, the physical origin of these features
is mainly related to the forward scattering of the probe wave $\Omega_1$ on
the collective spatiotemporal oscillations of the population difference
component $D_0(t, z)$. It is the transient ringing in the pump field
$\Omega_0$, that is able to modulate $D_0(t, z)$ [cf. Eq.~(\ref{4c})]. The
near-resonant features described are determined by the frequencies of
parametric modulation. Moreover, the importance of a value of the phase shift
between probe and pump fields, which is introduced by the time delay
$\tau_0$, substantiates the parametric treatment of the phenomenon.

Under the increase of the time delay between the probe and pump pulses
$|\tau_0|$, the efficiency of modifications in the probe spectrum decreases.
Under the decrease of the pump area $s_0$, the amplitude of the spectral
features also decreases, but its characteristic frequencies do not
significantly vary. The latter fact is different from the case of the
strong-field stationary Mollow-Boyd contour \cite{boyd}, where characteristic
frequencies depend on the amplitude of the pump field, but not on the length
and density of a medium.

\begin{figure}
\includegraphics{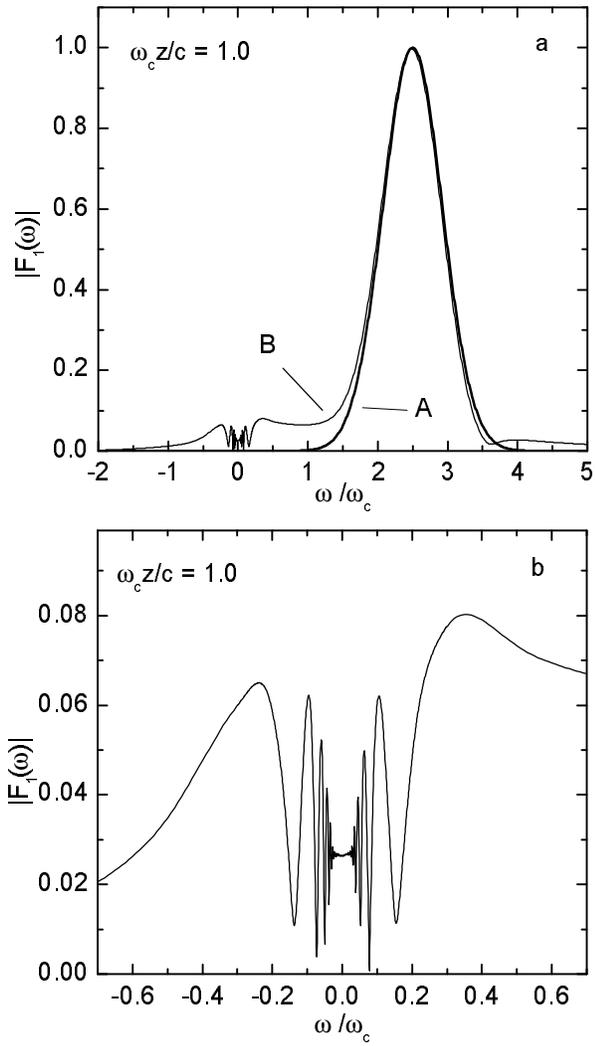}
\caption{\label{fig7}Spectra of the narrow-band probe field for propagation
distance $\omega_c z/c=1.0$. (a) input (curve A) and output (curve B)
spectra; (b) detailed output spectrum of (a) in the vicinity of resonance;
$\gamma_{1,2}/\omega_c=10^{-3}$, FWHM of the probe input-spectrum
$\gamma_\text{sp1} / \omega_c =1.0$, FWHM of the pump input-spectrum
$\gamma_\text{sp0} / \omega_c = 10$, pump area $s_0=0.49\pi$.}
\end{figure}

\begin{figure}
\includegraphics{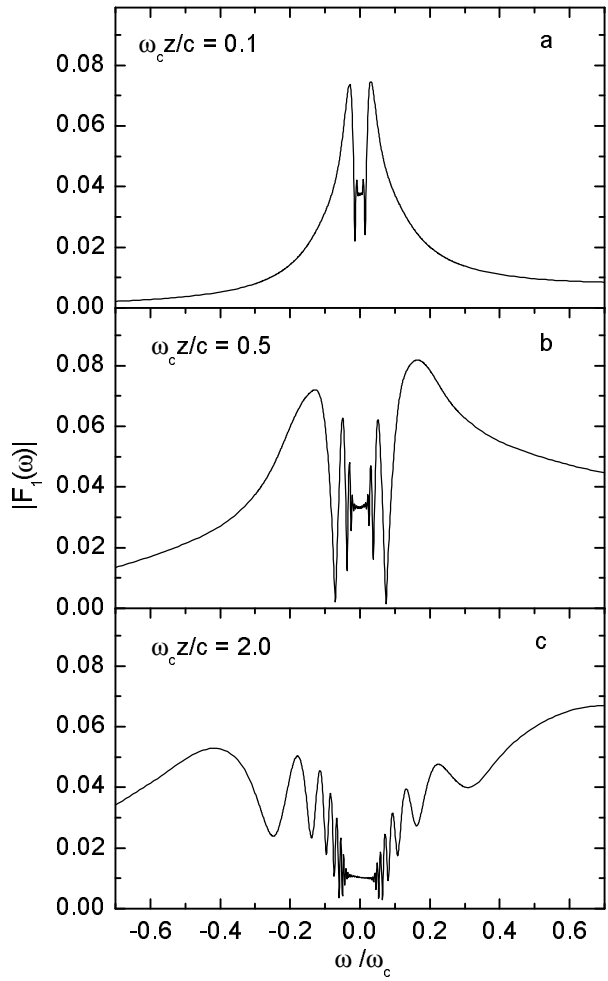}
\caption{\label{fig8}Output spectra of the narrow-band probe field in the
vicinity of resonance for different propagation distances. (a) $\omega_c
z/c=0.1$; (b) $\omega_c z/c=0.5$; (c) $\omega_c z/c=2.0$; corresponding input
spectrum is shown in Fig.~\ref{fig7}(a); $\gamma_{1,2}/\omega_c=10^{-3}$,
FWHM of the probe input-spectra $\gamma_\text{sp1} / \omega_c =1.0$, FWHM of
the pump input-spectra $\gamma_\text{sp0} / \omega_c = 10$, pump area
$s_0=0.49\pi$.}
\end{figure}

Parametric peculiarities of the process are also illustrated by the numerical
simulations with the narrow-band probe pulse shifted from the resonance. The
pump parameters were chosen to be the same as described above. The parameters
of the probe pulse had the following values: $\gamma_\text{sp1}/
\omega_c=1.0$, $\Delta_1/\omega_c=2.5$, and $\tau_0=0$. The results of the
simulations are presented in Figs.~\ref{fig7} and~\ref{fig8}.

Figure \ref{fig7} displays the input and output spectra of the narrow-band
probe pulse for the case of the propagation distance $\omega_c z/c=1.0$. The
figure shows that the characteristic doublet-like feature in the vicinity of
the resonance appears even in the case, where significant near-resonant
components are absent in the input spectrum. Therefore, the probe output
spectrum does not originate only from the amplification of existing
components, but also from the generation of new ones via modulation of medium
parameters by the pump field. Moreover, one can see the appearance of a
specific oscillating structure in the output spectrum.

The specific structure becomes richer and broader under the increase of the
propagation distance, which is demonstrated by Fig.~\ref{fig8} for $\omega_c
z/c=0.1$, 0.5, and 2.0 and by Fig.~\ref{fig7}(b) for  $\omega_c z/c=1.0$. The
total width of the amplification region also increases. The spectral
structure can be also seen in Figs.~\ref{fig4}(b) and~\ref{fig5} for the
broadband resonant probe field, though being stronger suppressed by the
absorption.

Such oscillating structure, which depends on the density and length of the
medium, is characteristic for spectral measurements in the presence of the
optical ringing. It was also obtained numerically \cite{mikl,schupper} and
experimentally \cite{ranka} under the investigation of the spectral
properties of self-induced transparency of single ultrashort pulses. The
structure was shown to appear if both nonlinear solution (particularly,
solitons) and optical ringing are significant in the output signal. The
experiments \cite{ranka} were carried out in a potassium vapor cell (the
interaction length was 2~mm) at D1 spectral line; the pulse duration was
equal to 415~fs, the vapor density reached a value of $10^{14}$~cm$^{-3}$,
which corresponds to the cooperative frequency $\omega_c/2\pi$ of about
90~GHz. The width of the oscillating structure obtained in this experiments
was about 40~GHz.

This kind of spectral features enables the analysis of the free-space optical
ringing by spectral methods. The features can be obtained even in the case of
linear interference with participation of the optical ringing in an optically
dense resonant medium. Particularly in the case, where a short laser pulse,
which have passed through a medium, interferes at the output with a similar
pulse, which have travelled the same distance in vacuum and having additional
arbitrary time delay $\tau_1$. Using the integrand of Eq.~(\ref{8}) and the
expression for the wave vector $k(\omega)$ (\ref{9}) in the limit of
$\gamma_2=0$, the summed spectrum of two fields $F_+(\omega)$ can be written
as follows:

\begin{equation}\nonumber
F_+(\omega)=F_\text{in}(\omega)e^{-i\omega z/c +i\omega_c^2z/(\omega c)} +
F_\text{in}(\omega)e^{-i\omega z/c -i\omega\tau_1}.
\end{equation}
Then, the absolute value of the spectrum is given by

\begin{equation}\label{19}
|F_+(\omega)|=2|F_\text{in}(\omega)|\left|\cos\left(\frac{1}{2}
\left(\frac{\omega_c^2 z}{\omega c} + \omega\tau_1\right)\right)\right| .
\end{equation}
In the region of the parameters, where the phase brought by the medium (the
first term) is greater than the phase brought by the time delay (the second
term), Eq.~(\ref{19}) gives the spectral oscillations, which are
qualitatively similar to the ones presented in Figs.~\ref{fig7}
and~\ref{fig8}. This phase is also responsible for the appearance of the
optical ringing. The function (\ref{19}) has an infinite number of extrema,
whose frequency increases when the detuning approaches the zero. If the
relaxation is taken into account, the number of oscillations is limited, and
the oscillations are absent in the region of the absorption.

So, the characteristic spectral oscillations can be made visible in the
spectrum of the optical ringing by the change of the phase in the linear
solution (\ref{8}), particularly, due to the presence of another pulse.
Moreover, it is the nonlinearity of the medium that can bring a specific
phase shift to the interaction process. In Refs.~\cite{mikl,ranka,schupper},
the appearance of the spectral oscillations can be treated as interference of
the quasilinear optical ringing with nonlinar solutions (particularly,
solitons), which have specific time delays due to the nonlinearity of
interaction. In the case of pump-probe interaction presented here, it is the
pump pulse that changes the propagation conditions of the probe. Unlike the
works \cite{mikl,ranka,schupper}, in our study, the spectral structure is
obtained in the weak amplified probe field intersected with the pump. The
input pump-pulse may have the area less than $\pi/2$. Moreover, contrary to
Refs.~\cite{mikl,ranka,schupper}, the amplitude of the input spectrum may be
negligibly small in the vicinity of the resonance and hence produces no
spectral background for the observation of the specific output features.

Thus, the pulsed pump-probe interaction may be significantly affected by the
generation of the optical ringing signal, and it is the dynamic peculiarities
of the collective reemission of a resonant medium that play a key role in the
processes considered. The efficiency of collective dynamics increases with
the increase of the propagation distance and density of a medium. On the
other hand, the efficiency of two-beam interaction may be limited by those
parameters, since under the increase of the oscillation frequency, the
amplitude of the field and population difference modulation decreases. This
fact reduces parametric coupling of two fields and leads to their linear
propagation.

\section{Conclusions}

We have presented the theoretical investigation of transient pump-probe
interaction between short laser pulses in an optically dense resonant medium.
A high value of the field-matter coupling coefficient, which is determined by
the medium density, and relatively weak intensities of the external pulses
give rise to the collective oscillating response of a medium in the form of
the coherent optical ringing. The long ringing response, which exists in the
time-scale of the order of relaxation times, provides the interaction of
light fields with a medium to be much longer than the duration of short input
pulses. The analytical expression obtained gives the relation between the
temporal behavior of the ringing and the spectrum of the input laser-pulse.
It describes the limitation of oscillation-frequency growth and the reduction
of the ringing efficiency under the increasing detuning from the resonance.
Moreover, it supports the treatment of the optical ringing as successive
coherent beats between light wave-packets of different group velocities,
which in the case of a dense coherent medium, can significantly vary over the
broad spectrum of the pulse.

The collective dynamics of the medium is shown to essentially affect the
propagation and amplification of a probe pulse under its nonlinear
interaction with the pump field. In the transmission spectrum, either
characteristic doublet-like structure or broad dip appears in the vicinity of
the resonant frequency, depending on the time delay between the probe and
pump pulses. In accordance with the spatiotemporal behavior of the optical
ringing, the widths of the coherent spectral features increase with the
increase of the medium density. Besides that, broadening of the coherent
doublet and dip during the propagation reflects the fact, that wave-packets
of increasing detunings from the resonance being involved in the interaction
of two fields. The modification of the spectrum comes from the modulation of
the probe and medium parameters by the pump pulse, especially, from the
oscillations of the population difference due to the existence of the optical
ringing in the pump field. Because of the modulation, the spectral features
can appear not only in the spectral region of the existing input components,
but also in the region, where the spectrum of the input probe-pulse is
negligibly small.

Thus, contrary to the stationary strong-field effects, the density- and
coordinate-dependent transmission spectra of the probe manifest the
importance of the collective oscillations and cannot be obtained in the
framework of the single-atom model. We would like to point out, that the
doublet spectral features and their sensitivity to the atomic density
(particularly, the existence of a threshold) qualitatively resemble the
analogous properties of the ``spectrum condensation'', or
``self-frequency-locking'', phenomenon \cite{vasilyev94,bertseva}.

\begin{acknowledgments}
The work was partially supported by the INTAS, Project No. 99-1366.
\end{acknowledgments}



\end{document}